\documentclass[graybox]{svmult}

\usepackage{mathptmx}       
\usepackage{helvet}         
\usepackage{courier}        
\usepackage{type1cm}        
\usepackage{makeidx}         
\usepackage{graphicx}        
\usepackage{multicol}        
\usepackage[bottom]{footmisc}

\makeindex             
                       
\usepackage{color}

\usepackage{epsfig,amssymb,psfrag,float,bm}
\usepackage[tbtags]{amsmath}

\usepackage{amsfonts,amsbsy,bm,euscript,mathrsfs}
\usepackage{amssymb,stmaryrd,faktor}
\usepackage{graphics}
\usepackage{tikz}
\usetikzlibrary{matrix,arrows}

\definecolor{hyperref}{RGB}{026,028,185}
\usepackage[bookmarks=true,colorlinks=true,linkcolor=hyperref,citecolor=hyperref,urlcolor=hyperref,bookmarksnumbered]{hyperref}

\def \be  {\begin{equation}}
\def \ee  {\end{equation}}
\def \ba  {\begin{eqnarray}}
\def \ea  {\end{eqnarray}}
\def\no{\nonumber}

\def\e{\epsilon}

\def\rmp{{\rm p}}

\newcommand{\arsinh}{\operatorname{arsinh}}

\newcommand{\IndA}{M}
\newcommand{\IndB}{N}
\newcommand{\IndC}{P}
\newcommand{\IndD}{Q}
\newcommand{\IndE}{R}
\newcommand{\IndF}{S}
\newcommand{\ms}{1}

\newcommand{\co}{\zeta}

\begin{document}

\title*{Unitarity methods for scattering in two dimensions}
\author{Valentina Forini, Lorenzo Bianchi and Ben Hoare}
\institute{Valentina Forini, Lorenzo Bianchi and Ben Hoare \at Humboldt Universit\"at zu Berlin, Newtonstrasse 15, 12489 Berlin, Germany\\ \email{forini,bianchi, hoare@physik.hu-berlin.de}
}
%
%
\maketitle

\abstract{The standard unitarity-cut method is applied to several massive two-dimensional models, including the world-sheet AdS$_5\times S^5$ superstring, to compute  $2\to 2$ scattering S-matrices at one loop from tree level amplitudes. Evidence is found for the cut-constructibility of supersymmetric integrable models, while for models without supersymmetry (but integrable) the missing rational terms can be interpreted as a shift in the coupling.}

\section{Discussion}

Unitarity-based methods, whose use in four dimensions has been crucial for an efficient evaluation of scattering amplitudes~\cite{Dixon:1996wi} in non-abelian gauge theories as well as gravity theories~\cite{ Elvang:2013cua}, have never really been applied in two dimensions~\footnote{For the three-dimensional case see~\cite{Chen:2011vv, CaronHuot:2012hr, Brandhuber:2013gda, Bianchi:2013pfa}.}. The aim of our work~\cite{Bianchi:2013nra} (we refer the reader to the independent results of~\cite{Engelund:2013fja}) has been to initiate the use of unitarity methods in the perturbative study of the S-matrix for \emph{massive two-dimensional} field theories. Limiting ourselves to the use of \emph{standard} unitarity (therefore placing on shell only two internal lines~\footnote{This is nothing but the application of the optical theorem. The case where the loop amplitude is subdivided into more than two pieces is referred to as \emph{generalized} unitarity.}) we present a formula for  the one-loop $2\to 2$ scattering amplitude built directly from the corresponding on-shell tree-level amplitudes. 

As reviewed below, we have applied our method to various models, finding enough evidence to postulate that \emph{supersymmetric, integrable two-dimensional theories should be cut-constructible via standard unitarity methods}. For bosonic theories with integrability,  we find agreement with perturbation theory up to a finite shift in the coupling~\footnote{It would be interesting to analyze models which are just  supersymmetric and not integrable.}. We also successfully apply our method to the light-cone gauge-fixed sigma-model for the $AdS_5\times S^5$  superstring, where - importantly - standard perturbation theory seems to fail  in evaluating the S-matrix beyond the leading order due to regularization issues.

Natural extensions of our analysis  would be the generalization to both higher loops~\footnote{Two-loop logarithmic contributions to the world-sheet scattering matrix for several backgrounds of interest were evaluated in~\cite{Engelund:2013fja}.} and higher points, as well as
the evaluation of rational contributions in the case of scattering of particles with different masses, interesting for example for the $AdS_3\times S^3\times M^4$ world-sheet S-matrix~\footnote{This implies an extension to the case of different masses of the t-channel prescription we describe in the next section, and would complete the analysis of~\cite{Engelund:2013fja} where the logarithmic part was computed up to two loops.}.

 
\section{Two-particle  S-matrix from unitarity cuts at one loop}
 
In two dimensions, the two-body scattering process of a translational-invariant field theory is described via the four-point amplitude
\begin{eqnarray}\nonumber 
&&\langle\Phi^\IndC(p_3)\Phi^\IndD(p_4)\,|\mathbb{S}|\,\Phi_\IndA(p_1)\Phi_\IndB(p_2)\rangle={\mathcal{A}}_{\IndA\IndB}^{\IndC\IndD}(p_1,p_2,p_3,p_4)\\\label{eqn:ampcons}
&&\qquad\qquad\equiv(2\pi)^2 \delta^{(2)}(p_1+p_2-p_3-p_4)\, \widetilde{\mathcal{A}}_{\IndA\IndB}^{\IndC\IndD}(p_1,p_2,p_3,p_4)~,
\end{eqnarray}
where $\mathbb{S}$ is the scattering operator, the fields $\Phi$ have on-shell momenta   $p_i$ (for us, all the particles  have equal non-vanishing mass  set to unity) and can carry flavor indices. Importantly, the energy-momentum conservation $\delta$-function satisfies
\be\label{delta2d}
\delta^{(2)} (p_1+p_2-p_3-p_4)=J(p_1,p_2)\,\big(\delta(\text{p}_1 - \text{p}_3)\delta(\text{p}_2 - \text{p}_4) +\delta(\text{p}_1 - \text{p}_4)\delta(\text{p}_2 - \text{p}_3) \big) \ ,
\ee
which accounts for the fact that in $d=2$ there is no phase space, and the only thing particles can do is either preserve or exchange their momenta. 
Above, $\rmp$ is the spatial momentum, the Jacobian $J(p_1,p_2)=1/(\partial \e_{\rmp_1}/\partial \rmp_1-\partial\e_{\rmp_2}/\partial \rmp_2)$ depends on the dispersion relation $\e_\rmp$  (the on-shell energy associated to $\rmp$) for the theory at hand, and spatial momenta are assumed to be ordered $\rmp_1>\rmp_2$.   
The S-matrix elements relevant for the description of the $2\to2$ scattering in the two-dimensional case are then defined~\footnote{Without loss of generality, one can consider in \eqref{eqn:ampcons} the amplitudes associated to the first product of $\delta$-functions $\delta(\text{p}_1 - \text{p}_3)\delta(\text{p}_2 - \text{p}_4)$. The denominator in \eqref{AandS} is required to make contact with the standard definition of the S-matrix in two dimensions.} as 
\be\label{AandS}
S_{\IndA\IndB}^{\IndC\IndD}(p_1,p_2)\equiv \frac{J(p_1,p_2)}{4\e_1 \e_2} \widetilde{\mathcal{A}}_{\IndA\IndB}^{\IndC\IndD}(p_1,p_2,p_1,p_2) ~.
\ee

\begin{figure}[ht]
\begin{center}
\begin{tikzpicture}[line width=2pt,scale=1.0]
\draw[-] (-5,-3) -- (-4,-4);
\draw[-] (-5,-5) -- (-4,-4);
\draw[-] (-2,-4) -- (-1,-3);
\draw[-] (-2,-4) -- (-1,-5);
\draw    (-3,-4) circle (1cm);
\draw[|-|,dashed,red!70,line width=1pt] (-3,-2.5) -- (-3,-5.5);
\draw[->] (-4.7,-3.0) -- (-4.3,-3.4);
\node at (-4.7,-2.8) {$p_1$};
\draw[->] (-4.7,-5.0) -- (-4.3,-4.6);
\node at (-4.7,-5.2) {$p_2$};
\draw[<-] (-1.3,-3.0) -- (-1.7,-3.4);
\node at (-1.3,-2.8) {$p_4$};
\draw[<-] (-1.3,-5.0) -- (-1.7,-4.6);
\node at (-1.3,-5.2) {$p_3$};
\draw[<-] (-3.1,-3.2) -- (-3.5,-3.35);
\node at (-3.2,-3.5) {$l_1$};
\draw[->] (-2.9,-4.8) -- (-2.5,-4.65);
\node at (-2.8,-4.5) {$l_2$};
\node at (-3.4,-2.85) {$\IndE$};
\node at (-2.6,-5.15) {$\IndF$};
\node at (-5.2,-3) {$\IndA$};
\node at (-5.2,-5) {$\IndB$};
\node at (-0.8,-5) {$\IndC$};
\node at (-0.8,-3) {$\IndD$};
\node [circle,draw=black!100,fill=black!5,thick,opacity=.95] at (-4,-4) {\footnotesize$\mathcal{A}^{(0)}$\normalsize};
\node [circle,draw=black!100,fill=black!5,thick,opacity=.95] at (-2,-4) {\footnotesize$\mathcal{A}^{(0)}$\normalsize};
\end{tikzpicture}
\\
\begin{tikzpicture}[line width=2pt,scale=1.0,rotate=90]
\draw[-] (-5,-3) -- (-4,-4);
\draw[-] (-5,-5) -- (-4,-4);
\draw[-] (-2,-4) -- (-1,-3);
\draw[-] (-2,-4) -- (-1,-5);
\draw    (-3,-4) circle (1cm);
\draw[|-|,dashed,red!70,line width=1pt] (-3,-2.5) -- (-3,-5.5);
\draw[->] (-4.7,-3.0) -- (-4.3,-3.4);
\node at (-4.7,-2.8) {$p_2$};
\draw[<-] (-4.7,-5.0) -- (-4.3,-4.6);
\node at (-4.7,-5.2) {$p_4$};
\draw[->] (-1.3,-3.0) -- (-1.7,-3.4);
\node at (-1.3,-2.8) {$p_1$};
\draw[<-] (-1.3,-5.0) -- (-1.7,-4.6);
\node at (-1.3,-5.2) {$p_3$};
\draw[<-] (-3.1,-3.2) -- (-3.5,-3.35);
\node at (-3.2,-3.5) {$l_1$};
\draw[<-] (-2.9,-4.8) -- (-2.5,-4.65);
\node at (-2.8,-4.5) {$l_2$};
\node at (-3.4,-2.85) {$\IndE$};
\node at (-2.6,-5.15) {$\IndF$};
\node at (-5.2,-3) {$\IndB$};
\node at (-5.2,-5) {$\IndD$};
\node at (-0.8,-5) {$\IndC$};
\node at (-0.8,-3) {$\IndA$};
\node [circle,draw=black!100,fill=black!5,thick,opacity=.95] at (-4,-4) {\footnotesize$\mathcal{A}^{(0)}$\normalsize};
\node [circle,draw=black!100,fill=black!5,thick,opacity=.95] at (-2,-4) {\footnotesize$\mathcal{A}^{(0)}$\normalsize};
\end{tikzpicture}
\hspace{10pt}
\begin{tikzpicture}[line width=2pt,scale=1.0,rotate=90]
\draw[-] (-5,-3) -- (-4,-4);
\draw[-] (-5,-5) -- (-4,-4);
\draw[-] (-2,-4) -- (-1,-3);
\draw[-] (-2,-4) -- (-1,-5);
\draw    (-3,-4) circle (1cm);
\draw[|-|,dashed,red!70,line width=1pt] (-3,-2.5) -- (-3,-5.5);
\draw[->] (-4.7,-3.0) -- (-4.3,-3.4);
\node at (-4.7,-2.8) {$p_2$};
\draw[<-] (-4.7,-5.0) -- (-4.3,-4.6);
\node at (-4.7,-5.2) {$p_3$};
\draw[->] (-1.3,-3.0) -- (-1.7,-3.4);
\node at (-1.3,-2.8) {$p_1$};
\draw[<-] (-1.3,-5.0) -- (-1.7,-4.6);
\node at (-1.3,-5.2) {$p_4$};
\draw[<-] (-3.1,-3.2) -- (-3.5,-3.35);
\node at (-3.2,-3.5) {$l_1$};
\draw[<-] (-2.9,-4.8) -- (-2.5,-4.65);
\node at (-2.8,-4.5) {$l_2$};
\node at (-3.4,-2.85) {$\IndE$};
\node at (-2.6,-5.15) {$\IndF$};
\node at (-5.2,-3) {$\IndB$};
\node at (-5.2,-5) {$\IndC$};
\node at (-0.8,-5) {$\IndD$};
\node at (-0.8,-3) {$\IndA$};
\node [circle,draw=black!100,fill=black!5,thick,opacity=.95] at (-4,-4) {\footnotesize$\mathcal{A}^{(0)}$\normalsize};
\node [circle,draw=black!100,fill=black!5,thick,opacity=.95] at (-2,-4) {\footnotesize$\mathcal{A}^{(0)}$\normalsize};
\end{tikzpicture}
\caption{Diagrams representing s-, t- and u-channel cuts contributing to the four-point one-loop amplitude.}
\label{stu}\nonumber
\end{center}
\end{figure}
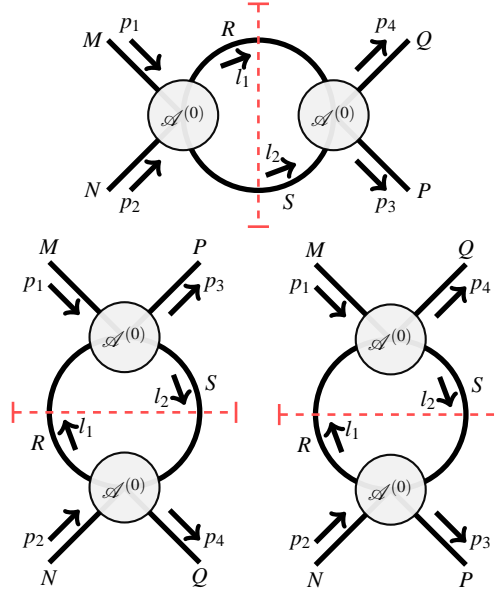

In applying the standard unitarity rules (derived from the optical theorem) \cite{Bern:1994zx} to the one-loop four point amplitude (\ref{eqn:ampcons}) one considers \emph{two-particle cuts}, obtained by putting two intermediate lines on-shell. The contributions that follow to the imaginary part of the amplitude are therefore given by the sum of $s$- $t$- and $u$-  channel cuts illustrated in Fig.~\ref{stu},  explicitly 
\ba\no
\mathcal{A}^{(1)}{}^{\IndC\IndD}_{\IndA\IndB}(p_1,p_2,p_3,p_4)|_{s-cut}=\int\frac{d^2 l_1}{(2\pi)^2}\int\frac{d^2 l_2}{(2\pi)^2}\ i\pi\delta^+({l_1}^2-\ms)\ i\pi\delta^+(l_2^2-\ms)\\\label{eqn:sch1}
\times\,\mathcal{A}^{(0)}{}_{\IndA\IndB}^{\IndE\IndF}({p_1,p_2,l_1,l_2})\mathcal{A}^{(0)}{}_{\IndF\IndE}^{\IndC\IndD} ({l_2,l_1,p_3,p_4})
\\\no
\mathcal{A}^{(1)}{}^{\IndC\IndD}_{\IndA\IndB}(p_1,p_2,p_3,p_4)|_{t-cut}=\int\frac{d^2 l_1}{(2\pi)^2}\int\frac{d^2 l_2}{(2\pi)^2}\ i\pi\delta^+({l_1}^2-\ms)\ i\pi\delta^+({l_2}^2-\ms)\\\label{eqn:tch1}
\times\,\mathcal{A}^{(0)}{}_{\IndA\IndE}^{\IndF\IndC}({p_1,l_1,l_2,p_3})\mathcal{A}^{(0)}{}_{\IndF\IndB}^{\IndE\IndD}({l_2,p_2,l_1,p_4})\\\no
\mathcal{A}^{(1)}{}^{\IndC\IndD}_{\IndA\IndB}(p_1,p_2,p_3,p_4)|_{u-cut}=\int\frac{d^2 l_1}{(2\pi)^2}\int\frac{d^2 l_2}{(2\pi)^2}\ i\pi\delta^+({l_1}^2-\ms)\ i\pi\delta^+({l_2}^2-\ms)\\\label{eqn:uch1}
\times\,\mathcal{A}^{(0)}{}_{\IndA\IndE}^{\IndF\IndD}({p_1,l_1,l_2,p_4})\mathcal{A}^{(0)}{}_{\IndF\IndB}^{\IndE\IndC}({l_2,p_2,l_1,p_3})
\ea
where $\mathcal{A}^{(0)}$ are tree-level amplitudes and a sum over the complete set of intermediate states $\IndE,\IndF$ (all allowed particles for the cut lines) is understood.  
Notice that tadpole graphs,  having no physical two-particle cuts,  are by definition ignored in this procedure.

To proceed, in each case one uses (\ref{eqn:ampcons}) and the  momentum conservation at the vertex involving the momentum $p_1$ to integrate over $l_2$, e.g. for the $s$-channel
\ba
 &&\!\!\!\!\!\!\!\!\!\!\!\!\!\!\!\!\!\!\!\!\!\!\!
 \mathcal{\widetilde{A}}^{(1)}{}^{\IndC\IndD}_{\IndA\IndB}(p_1,p_2,p_3,p_4)|_{s-cut}=\int\frac{d^2 l_1}{(2\pi)^2}\,i\pi\delta^+({l_1}^2-\ms)\, i\pi\delta^+(({l_1}-{p_1}-{p_2})^2-\ms)\nonumber\\
&&\qquad\times \,\,\widetilde{\mathcal{A}}^{(0)}{}_{\IndA\IndB}^{\IndE\IndF}({p_1,p_2,l_1,-l_1+p_1+p_2}) \,\widetilde{\mathcal{A}}^{(0)}{}_{\IndF\IndE}^{\IndC\IndD}({-l_1+p_1+p_2,l_1,p_3,p_4})\,,\label{2_8} 
\ea
The simplicity of the two-dimensional kinematics and of being at one loop plays now its role, since in each of the  integrals the set of zeroes of the $\delta$-functions is a discrete set, and the cut loop-momenta are frozen to specific values~\footnote{ At two loops, to constrain completely the four components of the two momenta circulating in the loops  one needs four cuts, each one giving an on-shell $\delta$-function. Two-particle cuts at two loops would result in a manifold of conditions for the loop momenta.}. This allows us to pull out the tree-level amplitudes with the loop-momenta evaluated at those zeroes~\footnote{This is like using $f(x)\delta(x-x_0)=f(x_0)\delta(x-x_0)$, where $f(x)$ are the tree-level amplitudes in the integrals.}. In what remains, following standard unitarity computations \cite{Bern:1994zx}, we apply the replacement $i\pi \delta^+(l^2-1) \longrightarrow \tfrac{1}{l^2-1}$ (\emph{i.e.} the Cutkowsky rule in reverse order) which sets loop momenta back off-shell, thus reconstructing scalar bubbles.  
 This allows us to rebuild, from its imaginary part, the cut-constructible piece of the amplitude and, via \eqref{AandS}~\footnote{This corresponds to the choice $p_3=p_1$, $p_4=p_2$.}, of the  S-matrix.  It then follows that a candidate expression for the
one-loop S-matrix elements is given by the following simple sum of products of two tree-level amplitudes~\footnote{In \eqref{eqn:final}, $\tilde S^{(0)}(p_1,p_2)=4 (\e_2\,\rmp_1-\e_1\,\rmp_2) S^{(0)}(p_1,p_2)$ and the denominator on the right-hand side comes from the Jacobian $J(p_1,p_2)$ assuming a standard relativistic dispersion relation (for the theories we consider, at one-loop this is indeed the case). } 
\begin{eqnarray} \label{eqn:final}
&& {S^{(1)}}{}_{\IndA\IndB}^{\IndC\IndD}(p_1,p_2)=\frac{1}{4 (\e_2\,\rmp_1-\e_1\,\rmp_2)}\,\Big[ {\tilde S^{(0)}}{}_{\IndA\IndB}^{\IndE\IndF}(p_1,p_2){\tilde S^{(0)}}{}_{\IndE\IndF}^{\IndC\IndD}(p_1,p_2)\,I_{p_1+p_2}\\ \nonumber\vphantom{\frac{1}{4 (\e_2\,\rmp_1-\e_1\,\rmp_2)}}
 &&+ {\tilde S^{(0)}}{}_{\IndA\IndE}^{\IndF\IndC}(p_1,p_1){\tilde S^{(0)}}{}_{\IndF\IndB}^{\IndE\IndD}(p_1,p_2)\,I_0 
 \vphantom{\frac{1}{4 (\e_2\,\rmp_1-\e_1\,\rmp_2)}}
 +{\tilde S^{(0)}}{}_{\IndA\IndE}^{\IndF\IndD}(p_1,p_2){\tilde S^{(0)}}{}_{\IndF\IndB}^{\IndC\IndE}(p_1,p_2)\,I_{p_1-p_2}\,\Big]
\end{eqnarray}
where the coefficients are given in terms of the bubble integral
\be
I_p=\int \frac{d^2 q}{(2\pi)^2} \frac{1}{(q^2-\ms+i\e) ((q-p)^2-\ms+i\e)}~
\ee 
 and read explicitly 
\begin{eqnarray}\nonumber
 I_{p_1+p_2}=\frac{i\pi-\arsinh(\e_2\,\rmp_1-\e_1\,\rmp_2)}{4\pi i\,(\e_2\,\rmp_1-\e_1\,\rmp_2)},\qquad
 I_0=\frac{1}{4\pi i},\qquad
 I_{p_1-p_2}=\frac{\arsinh(\e_2\,\rmp_1-\e_1\,\rmp_2)}{4\pi i\,(\e_2\,\rmp_1-\e_1\,\rmp_2)}~. 
\end{eqnarray}
A few importants remarks are in order:
\begin{description}
\item (a) Since the unitarity-cut procedure only ensures the correctness of logarithmic terms (in general, of those  terms associated to branch-cut singularities, typically logarithms or polylogarithms), the proposal \eqref{eqn:final} and its fermionic generalization~\cite{Bianchi:2013nra}  crucially need to be tested on known examples~\footnote{Because its bubble integral $I_0$ can only contribute to rational terms, the $t$-channel contribution has been neglected in~\cite{Engelund:2013fja}, where all rational terms were determined from symmetry considerations.}.  
\item (b) The $t$-channel cut requires a prescription, since if one first uses the $\delta$-function identity \eqref{delta2d} to fix, for example, $p_1 = p_3$ and $p_2 = p_4$ the corresponding integral is ill-defined.  To avoid this ambiguity we follow the prescription that we should only impose the $\delta$-function identity  at the end~\footnote{In some sense this is natural as, in general dimensions, quantum field theory amplitudes have the form \eqref{eqn:ampcons}, while the $\delta$-function identity \eqref{delta2d} is specific to two dimensions.}. Furthermore, if we choose the alternative solution of the conservation $\delta$-function in (\ref{eqn:tch1}), namely $\ell_2=\ell_1+p_4-p_2$, the coefficient of $I(0)$ in (\ref{eqn:final}) would be different, which leads to the consistency condition on the tree-level S-matrix~\footnote{See \cite{Bianchi:2013nra} for the  generalization to the case which includes fermions.} 
\be\label{consistency}
{\tilde S^{(0)}}{}_{\IndA\IndE}^{\IndF\IndC}(p_1,p_1)\,{\tilde S^{(0)}}{}^{\IndE\IndD}_{\IndF\IndB}(p_1,p_2)\,=\,{\tilde S^{(0)}}{}^{\IndC\IndF}_{\IndA\IndE}(p_1,p_2)\,{\tilde S^{(0)}}{}^{\IndD\IndE}_{\IndF\IndB}(p_2,p_2)~.
\ee
We have checked this for the tree-level S-matrices of all the field theory models treated below.
\item (c) As they only involve the scalar bubble integral in two dimensions,  the result~\eqref{eqn:final} following from our procedure is inherently \emph{finite}. No additional regularization is required and the result can be compared directly with the $2 \to 2$ particle S-matrix (following from the finite or renormalized four-point amplitude) found using standard perturbation theory.
Of course, this need not be the case for the original bubble integrals before cutting -- due to factors of loop-momentum in the numerators. These divergences, along with those coming from tadpole graphs, which we did not consider, should be taken into account for the renormalization of the theory. We have not investigated this issue, since all the theories we consider below   are either UV-finite or renormalizable. 
\end{description}
To explore the validity of the procedure outlined we have considered both relativistic and non-relativistic (world-sheet field theory for the AdS$_5\times S^5$ superstring) models. 
 
 \section{Relativistic models }

In the  \emph{relativistic, bosonic case},  we looked at a class of generalized sine-Gordon models \cite{Hollowood:1994vx,Bakas:1995bm}, theories defined by a gauged WZW model for a coset $G/H$ plus a potential, whose classical integrability can be demonstrated through the existence of a Lax connection. Considering the coset $G/H = \text{SO}(n+1)/\text{SO}(n)$, where asymptotic excitations are a free $\text{SO}(n)$ vector with unit mass (which is the case we have considered in our general procedure), this class includes the sine-Gordon ($n=1$) and complex sine-Gordon ($n=2$) models,  for which the exact S-matrices are known \cite{Zamolodchikov:1978xm,Dorey:1994mg}. In all the cases the one-loop S-matrix got via unitarity cuts agrees -- up to a term proportional to the tree-level S-matrix which can be interpreted as a scheme-dependent shift in the coupling~\footnote{In the sine-Gordon case the agreement is exact. For $n\geq 2$ the shift in the coupling is by the dual Coxeter number of the group $G=\text{SO}(n)$, a structure that appears regularly in the quantization of WZW and gauged WZW models, where $k$ is the quantized level (see for example~\cite{Witten:1983ar}).} --  with the one known from perturbation theory. Importantly, the latter includes one-loop corrections coming from a gauge-fixing procedure which integrates out unphysical fields \cite{Hoare:2010fb} and results in contributions to the one-loop S-matrix which restore various properties of integrability. 

As for  \emph{relativistic, supersymmetric models},  which have checked the procedure on theories obtained as Pohlmeyer reductions of the  Green-Schwarz action for the Type IIB superstring on AdS$_5 \times S^5$\cite{Grigoriev:2007bu,Mikhailov:2007xr},  AdS$_3 \times S^3$ \cite{Grigoriev:2008jq} and AdS$_2 \times S^2$ \cite{Grigoriev:2007bu}  which can be seen as supersymmetric generalizations of the bosonic models considered above~\footnote{The reduced AdS$_2 \times S^2$ theory is in fact given by the $\mathcal{N} = 2$ supersymmetric sine-Gordon model. The reduced AdS$_3 \times S^3$ and AdS$_5 \times S^5$ theories have a non-local $\mathcal{N}=4$ and $\mathcal{N}=8$ supersymmetry respectively, which manifests as a $q$-deformation of the S-matrix symmetry algebra. We have also checked that the unitarity-cutting procedure matches the perturbative result at one-loop in the $\mathcal{N}=1$ supersymmetric sine-Gordon model \cite{Shankar:1977cm}.}.  These reduced theories are all classically integrable, demonstrated by the existence of a Lax connection, and conjectured to be UV-finite \cite{Roiban:2009vh}. 
 The tree-level and one-loop S-matrices for these theories were computed in \cite{Hoare:2009fs,Hoare:2011fj}, while the exact S-matrices have been conjectured using integrability techniques in \cite{Kobayashi:1991rh} for the reduced AdS$_2 \times S^2$ model, \cite{Hoare:2011fj} for the reduced AdS$_3 \times S^3$ model and \cite{Hoare:2013ysa} for the reduced AdS$_5 \times S^5$ model.  In all the cases considered, the agreement is \emph{exact} and no additional shift of the coupling is needed.  The presence of the supersymmetry, albeit deformed, may provide an explanation for this, with shifts arising from bosonic loops cancelled by shifts from fermionic loops.   
Importantly,  in the reduced AdS$_3 \times S^3$ standard perturbative computation a contribution coming from a one-loop correction needs to be added so that the S-matrix satisfies the Yang-Baxter equation. It is this S-matrix that the unitarity technique matches. 
This is then another example of how unitarity methods applied to a classically integrable theory seem to provide a \emph{quantum integrable} result. This seems to suggest a relationship between integrable quantization and unitarity techniques which would be interesting to investigate further. 

\section{ AdS$_5\times S^5$ superstring world-sheet theory}

We have finally considered the  case of the light-cone gauge-fixed superstring on AdS$_5 \times S^5$ and its world-sheet S-matrix~\footnote{Notice that this is a \emph{non-relativistic} model, as seen quantizing it perturbatively and noticing that the choice of a flat Minkowski worldsheet metric is incompatible with Virasoro constraints (see for example~\cite{Callan:2003xr}).}.
 Assuming the quantum integrability of the full world-sheet theory and using the global symmetries the \emph{exact} world-sheet S-matrix has been uniquely determined~\cite{Beisert:2005tm} up to an overall phase, or dressing factor~\cite{Arutyunov:2004vx}. The determination of the latter exploited the non-relativistic generalization of the crossing symmetry~\cite{Janik:2006dc,Volin:2009uv}
as well as perturbative data both from the string and gauge theory sides~\cite{Beisert:2006ib,Beisert:2006ez}. 
Relaxing the level-matching condition and taking the limit of infinite light-cone momentum (decompactification limit), the world-sheet theory becomes a massive field theory defined on a plane, with well-defined asymptotic states and S-matrix. 
The scattering of the world-sheet excitations has been studied at tree-level in~\cite{Klose:2006zd}, while one-loop~\cite{Klose:2007wq} and two-loop~\cite{Klose:2007rz} results have been carried out only in the simpler near-flat-space limit~\cite{Maldacena:2006rv} where interactions are at most quartic in the fields. These studies have also explicitly shown some consequences of the integrability of the model, such as the factorization of the many-body S-matrix  and the absence of particle production in the scattering processes~\cite{Puletti:2007hq}.

The tree level matrix elements were evaluated in~\cite{Klose:2006zd}  in the generalized uniform light-cone gauge (showing therefore  an explicit dependence on the parameter $a$ labeling different light-cone gauge choices~\cite{Arutyunov:2006gs}) at leading order in perturbation theory, where the small parameter  
is the inverse of the string tension $ g = \frac{\sqrt{\lambda}}{2\pi} \ .$
After having explicitly verified that the tree-level matrix elements above verify the fermionic generalization of the consistency relation \eqref{consistency}, we could safely use them as an input of our procedure  and get the one-loop S-matrix for the  light-cone gauge-fixed sigma model~\footnote{Notice that the non-relativistic dispersion relation $\epsilon(\rmp)=\sqrt{1+\frac{\lambda}{\pi^2}\sin^2\frac{\rmp}{2}}$~\cite{Beisert:2004hm,Beisert:2005tm}, when expanded in the near-BMN limit $\rmp \to \co \rmp$, corresponding to the perturbative regime, leads to a relativistic energy $\e_i=\sqrt{1+\rmp_i^2}$.}. As a first result,  an overall phase could be resummed at the one-loop order, which show  the expected gauge dependence~\cite{Arutyunov:2009ga}.
 As mentioned above, because of the complicated structure of interactions of the light-cone gauge-fixed sigma model,
the perturbative S-matrix is known beyond the leading order~\cite{Klose:2007wq,Klose:2007rz} only in the   kinematic truncation 
known as near-flat-space limit~\cite{Maldacena:2006rv}.
Therefore, to test the validity of the unitarity method,  we needed to compare our one-loop result to the corresponding limit of the exact world-sheet S-matrix. 
This was achieved by considering the matrix elements derived in~\cite{Beisert:2005tm} for a single SU$(2|2)$ sector together with the dressing phase, here needed at next-to-leading order in the $1/\sqrt{\lambda}$ expansion~\footnote{In the comparison with the world-sheet calculation all dimensional quantities (such as the spin-chain length and the momenta) should be rescaled via a factor of $\sqrt{\lambda}/(2\pi)$~\cite{Klose:2006zd}, for us $\rmp\to \co\,\rmp$.}. 

In comparing the exact S-matrix with the one found via unitarity cuts~\footnote{This is done in the so-called constant-$J$ gauge $a = 0$.}  we found
\begin{equation}\label{strings_compar_2}
({{S}}_{AB}^{CD})_{\rm exact}= 
e^{\frac{i}{4\,g} \big(([A] + 2[B] -[C] - 2) \rmp_1 +([B] - 2[C] - [D]+2) \rmp_2\big)} \,e^{\varphi_{a=0}(\rmp_1,\rmp_2)} \, (S_{AB}^{CD})_{\rm cut} + \mathcal{O}(1/g^3) \ . 
\end{equation}
From \eqref{strings_compar_2} we see that we have agreement up to a phase whose argument is linear in momenta.
This is not surprising, as it simply amounts to moving from the string frame to the spin-chain frame~\cite{Arutyunov:2006yd,Ahn:2010ka}. As argued already at the tree level~\cite{Klose:2006zd}, such terms should not
affect the physical spectrum following from inputting the S-matrix into the asymptotic Bethe equations.

\bibliographystyle{nb}

\bibliography{Ref_2dcuts}

\end{document}